\documentclass[lettersize,journal]{IEEEtran}
\IEEEoverridecommandlockouts
\usepackage{cite}
\usepackage{amsmath,amssymb,amsfonts}
\usepackage{algorithmic}
\usepackage{graphicx}
\usepackage{textcomp}
\usepackage{xcolor,url}
\usepackage{subcaption,multirow,cleveref}

\crefname{algocf}{alg.}{algs.}
\Crefname{algocf}{Algorithm}{Algorithms}

\usepackage[ruled,linesnumbered,vlined]{algorithm2e}
\newcommand{\nonl}{\renewcommand{\nl}{\let\nl\oldnl}}
\makeatletter
\newcommand{\nosemic}{\renewcommand{\@endalgocfline}{\relax}}
\newcommand\independent{\protect\mathpalette{\protect\independenT}{\perp}}
\def\independenT#1#2{\mathrel{\rlap{$#1#2$}\mkern2mu{#1#2}}}
\def\BibTeX{{\rm B\kern-.05em{\sc i\kern-.025em b}\kern-.08em
    T\kern-.1667em\lower.7ex\hbox{E}\kern-.125emX}}
\begin{document}


\title{Intrinsically Motivated Reinforcement Learning based Recommendation with Counterfactual Data Augmentation}

\author{Xiaocong Chen,~\IEEEmembership{Student Member, IEEE}, Siyu Wang, Lina Yao,~\IEEEmembership{Senior Member, IEEE}, Lianyong Qi,~\IEEEmembership{Member, IEEE}, Yong Li,~\IEEEmembership{Senior Member, IEEE}
\thanks{X. Chen, S. Wang, and L. Yao are with the School 
of Computer Science and Engineering, University of New South Wales, Sydney,
NSW, 2052, Australia. \protect E-mail: xiaocong.chen@unsw.edu.au}
\thanks{L.Yao is with CSIRO's Data61, Australia}
\thanks{L. Qi is College of Computer Science and Technology, China University of Petroleum (East China), China}
\thanks{Y, Li is with Dept. of Electronic Engineering, Tsinghua University, China, 100084}
}
        
\markboth{Journal of \LaTeX\ Class Files,~Vol.~14, No.~8, August~2021}%
{Shell \MakeLowercase{\textit{et al.}}: A Sample Article Using IEEEtran.cls for IEEE Journals}


\maketitle

\begin{abstract}
Deep reinforcement learning (DRL) has been proven its efficiency in capturing users' dynamic interests in recent literature. However, training a DRL agent is challenging, because of the sparse environment in recommender systems (RS), DRL agents could spend times either exploring informative user-item interaction trajectories or using existing trajectories for policy learning. It is also known as the exploration and exploitation trade-off which affects the recommendation performance significantly when the environment is sparse. It is more challenging to balance the exploration and exploitation in DRL RS where RS agent need to deeply explore the informative trajectories and exploit them efficiently in the context of recommender systems. As a step to address this issue, We design a novel intrinsically ,otivated reinforcement learning method to increase the capability of exploring informative interaction trajectories in the sparse environment, which are further enriched via a counterfactual augmentation strategy for more efficient exploitation. The extensive experiments on six offline datasets and three online simulation platforms demonstrate the superiority of our model to a set of existing state-of-the-art methods. 
\end{abstract}


\section{Introduction}
Recently, deep reinforcement learning (DRL) receives increasing interests in RS because of its capability in capturing users' dynamic interests~\cite{chen2021survey}. Current DRL-based RS can be generally categorized into three streams: value-based methods, policy-based methods and hybrid methods. One of the representatives of value-based methods would be the Deep Q-learning (DQN), in which~\cite{zheng2018drn} bring it into news recommendation. However, deep Q-learning based methods require the ``maximize'' operation over the action space (i.e., all the candidate items) which is not traceable and may induce agent stuck problem~\cite{dulac2015deep}. Policy-gradient methods can mitigate such problem but suffers the high variance problem as the optimization is based on last step's trajectory which could be distinct from previous trajectories~\cite{xu2020reinforcement}. Hybrid method is the combination of policy-gradient and value-based methods. It aims to reduce the variance for policy-gradient by introducing the value-based method~\cite{degris2012off} and gains more attention \cite{chen2020knowledge,chen2019large,cai2018reinforcement,chen2021generative}.

However, the user-item interactions are commonly sparse. It hinders policy optimisations to find the rewards via exploration as well as maximizing the performance via exploitation. Specifically, as DRL relies on carefully engineering environment rewards that are extrinsic to the agents, the sparsity barely provides dense reward signals (i.e., most of the reward signals might be missing because of the highly incomplete interactions and user feedback). 
Hence, new exploration strategies would be a choice to encourage agent to discover a wider range of states and formulate richer interaction trajectories~\cite{nair2018overcoming}. Recent literature~\cite{chen2021values,chen2021exploration} show that 
exploration is effective to reduce the model uncertainty in regions of sparse rewards or user interactions. Moreover, most of existing works in DRL RS apply $\epsilon$-greedy as the exploration strategy that agent has $\epsilon$ possibility to conduct exploration randomly~\cite{chen2021survey}. However, random exploration increases the training time and the uncertainty and may not be able to explore enough informative interaction trajectories. Moreover, it will cost considerable amount of trials and it is not feasible to be applied for highly sparse user feedback in the recommender systems as it also requires significant amount of trials for exploitation which is known as exploration and exploitation trade-off. 

Differently, several attempts have been made from the different perspective of data augmentation to relief the sparsity. Experience replay is widely used in DRL methods which empowers agent to learn by reusing past interaction trajectories. However, the experience replay can only promote certain trajectory to be replayed~\cite{schaul2015prioritized}. The policy learning process may be harmed if the generated trajectories are not informative. Recent studies also investigate to equip data augmentation with causality by governing to generate informative trajectories. For instance,~\cite{wang2021counterfactual} designs a simple counterfactual method by measuring the embedding changing to generate new user sequence. 

Moreover, \cite{zhang2021causerec} considers the embedding contains two parts which are dispensable or indispensable items related to the final recommended items by leveraging the causality. By replacing dispensable items, it can generate more user sequence but with the same performance. The main limitation with these approaches, however, is that they assume an embedding of state space, while state is dynamic and updated after each interaction. Moreover, the agent never knows the ground-truth (i.e., user's final choice) during the online interactions. Hence, it is impractical to leverage the ground-truth to determine the embedding difference or indispensable items as existing works did.

In order to address the above issues, we propose a new end to end model namely Intrinsically Motivated Reinforcement Learning with Counterfactual Augmentation (IMRL) from two aspects: augmenting informative trajectories and a new exploration strategy. We design an novel empowerment-based exploration strategy to encourage agent to explore the potentially informative interaction trajectories in the sparse environment. Moreover, we elaborate a new counterfactual data augmentation method for DRL RS to augment those newly explored informative trajectories so that they can have a higher exposure probability thus boost the final performance.

In summary, we make the following contributions in this paper:
\begin{itemize}
    \item We propose a novel DRL method IMRL which  can augment trajectories which is causally valid but never seen by the agent to relief the data sparsity problem. Moreover, we also introduce an adaptive threshold to dynamically control the boundary of the informative as the learning process in DRL is evolutionary.
    \item We design an empowerment-driven exploration strategy for IMRL to help explore those un-explored but potentially informative interaction trajectories. Our experiments show that the designed exploration strategy can boost the final performance in the online simulation platforms.
    \item We have conduct extensive experiments in both offline and online settings and show the superiority of IMRL. We have conduct offline experiments with six well-known datasets and online experiments in three public simulation platforms.
\end{itemize}

\section{Background}
\subsection{Problem Formulation}
Reinforcement learning based recommender systems learn from interactions through a Markov Decision Process (MDP).
Given a recommendation problem consisting of a set of users $\mathcal{U} = \{u_0,u_1,\cdots$ $u_n\}$, a set of items $\mathcal{I} = \{i_0,i_1,\cdots i_m\}$ and user's demographic information $\mathcal{D}=\{d_0,d_1,\cdots,d_n\}$, MDP can be represented as a tuple $(\mathcal{S},\mathcal{A},\mathcal{P},\mathcal{R},\gamma)$, where each represents the following,
\begin{itemize}
    \item $\mathcal{S}$ denotes the state space, which is the combination of the subsets of $\mathcal{I}$ and $\mathcal{D}$, it represents user's previous interactions and demographic information. Based on that, it can be written as a composition form: $\mathcal{S} = \mathcal{S}^1 \oplus \mathcal{S}^2 \oplus \cdots \mathcal{S}^n$ for a fixed $n$ which represents the dynamic count of components~\cite{zaheer2017deep};
    \item $\mathcal{A}$ is the action space, which represents agent's selection during recommendation based on the state space $\mathcal{S}$. Similarly, it can also be written as a composition form: $\mathcal{A} = \mathcal{A}^1 \oplus \mathcal{A}^2 \oplus \cdots \mathcal{A}^n$;
    \item $\mathcal{P}$ is the set of transition probabilities for state transfer based on the action received which also refers to users' behavior probabilities, it is worth to mention that $\mathcal{P}$ will not be estimated in this study as we are using model-free reinforcement learning approach;
    \item $\mathcal{R}$ is a set of rewards received from users, which are used to evaluate the action taken by the recommendation system, with each reward being a binary value to indicate user's click; $\gamma$ is a discount factor;
    \item $\gamma \in [0,1]$ for balancing the future reward and current reward.
\end{itemize}  

Given a user $u_0$ and the state $s_0$ observed by the agent (or the recommendation system), which includes a subset of item set $\mathcal{I}$ and user's demographic information $d_0$, a typical recommendation iteration for user $u_0$ goes as follows.
First, the agent makes an action $a_0$ based on the recommend policy $\pi_0$ under the observed initial state $s_0$ and receives the corresponding reward $r_0$.
Then, the agent generates a new policy $\pi_1$ based on the received reward $r_0$ and determines the new state $s_1$ based on the probability distribution $p(s_{new}|s_0,a_0)\in\mathcal{P}$.
The cumulative reward after $k$ iterations is as follows:
\begin{align*}
  r_c = \sum_{k=0} \gamma^{k}r_k
\end{align*}
\subsection{Local Causal Models}
Structural Causal Models (SCMs)~\cite{pearl2009causal} can be represented as a tuple: $\mathcal{M}_t(V_t,U_t,F)$
with the following components based on the state and action composition form at timestamp $t$. It normally represents as a directed acyclic graph (DAG) $\mathcal{G}$ with the following components,
\begin{itemize}
    \item $V_t = \{s_t^1, s_t^2, \cdots, s_t^n, a_t^1,\cdots,a_t^m, s_{t+1}^1,\cdots,s_{t+1}^n\}$ represents the nodes in DAG $\mathcal{G}$.
    \item $U_t = \{u^1, \cdots, u^{2n+m}\}$ is a set of noise variables, one for each in $V_t$. It is determined by the initial state, past actions and environment. We assume that the noise variable is time independent. Which implies that $U_t \independent U_{t+1}$.
    \item $F$ is a set of functions that: $U_t \times \text{Parentage}(V_t) \rightarrow V_t$ where $\text{Parentage}(\cdot)$ means the parent node of $\cdot$.
\end{itemize}
where we assume that the dynamic count of state is $n$ and $m$ for action, state observed at timestamp $t$ is written as $s_t$ which is the composition of $\{s_t^1, s_t^2, \cdots, s_t^n\}$.

Local causal models is an extension to the SCM which only considers about the local causal effect~\cite{pitis2020counterfactual}. Local causal model can be represented as $\mathcal{M}_t^\mathcal{L}(V_t^\mathcal{L},U_t^\mathcal{L},F^\mathcal{L})$ with DAG $\mathcal{G}^\mathcal{L}$ from the global causal model $\mathcal{M}_t$ in the subspace $\mathcal{L}$ with the same components with extra constraint,
\begin{align}
    & \text{Parentage}(V_t^\mathcal{L}) = \text{Parentage}(V_t|(s_t,a_t)\in\mathcal{L}), \\
    & \text{Parentage}(U_t^\mathcal{L}) = \text{Parentage}(U_t|(s_t,a_t)\in\mathcal{L}).
\end{align}
Moreover, the local causal model requires the set of edges in $\mathcal{G}$ to be structurally minimal~\cite{peters2017elements}. 

\section{Methodology}

\begin{figure}[h]
    \centering
    \includegraphics[width=\linewidth]{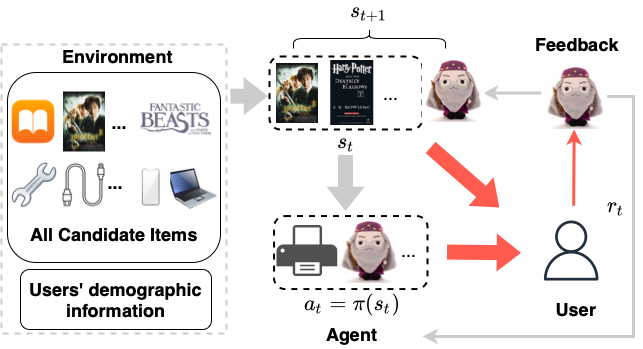}
    \caption{Illustration of a typical interaction between user and DRL agent in DRL RS. Red line represents user's information flow and grey represents receommender information flow. The states are sampled from the environment.}
    \label{fig:workflow}
\end{figure}

In this section, we will briefly explain the proposed approach Intrinsically Motivated Reinforcement Learning with Counterfactual Augmentation (IMRL) for reinforcement learning based recommendation which can address the sparse interactions problem in DRL RS. 
We are addressing such problem from two aspects according to the aforementioned challenges: i) use a novel data augmentation method to generate more potentially informative interaction trajectories by using using counterfactual reasoning; ii) design a new exploration strategy by introducing an intrinsic reward signal to encourage agent to conduct the exploration.

Hence, the proposed IMRL consists two main components: \textit{Counterfactual reasoning for data augmentation} and \textit{Intrinsically Motivated Exploration}.

\subsection{Counterfactual Data Augmentation}
Formally, given an arbitrary trajectory $\tau:(s,a,r,s')$ that sampled from the replay buffer, $r$ is the reward signals received by agent when action $a$ is executed at state $s$. Given the large candidate item set situation, most of the trajectories are not informative (i.e., $r$ is zero). With the zero reward situation, the informative trajectories are hard to be sampled because the number of those non-informative trajectories is significantly larger. Hence, augment those informative trajectories to increase the possibility to be sampled is a straightforward solution. 

We assume that state $s_{t+1}$ satisfy the SCM:
\begin{align}
    s_{t+1} = f(s_t,a_t,U_{t+1}),
\end{align}
where $f(\cdot)$ represents the causal mechanism, $a_t$ is the action at timestamp $t$, and $U_{t+1}$ is the noise term which is independent of $(s_t,a_t)$. Our main idea is to estimate the causal mechanism $f(\cdot)$ and generate more data that unseen but causally valid. To achieve the goal, the simple solution is to work out the causal mechanism $f(\cdot)$ firstly. However, estimate the global $f(\cdot)$ is challenging~\cite{lu2020sample}. Instead, inspired by recent advance in local causal models~\cite{xiang2013acquisition,pitis2020counterfactual}, we decide to estimate the local causal mechanism $f_l(\cdot)$. Local causal model assumes that there exists a local DAG $\mathcal{G}^{\mathcal{L}(i\independent j)}$ in a subspace $\mathcal{L}^{i\independent j}$ such that $\mathcal{L}^{i\independent j} \subset \mathcal{L}$ where $\mathcal{L}: \mathcal{S}\times\mathcal{A}$. It satisfies the following condition:
\begin{align}
    s_{t+1}^j \independent V_t^i | \text{Parentage}(s_{t+1}^j)\setminus V_t^i, (s_t,a_t) \in \mathcal{L}^{j\independent i},
\end{align}
Specifically, in recommender systems, there is a large subspace of states in which the users' previous interests will not affect the final recommendation as user's interest are dynamics. Hence, if we focus on the subspace $(s_t,a_t)\in\mathcal{L}^{(j\independent i)}$, we can formulate a local causal model $\mathcal{M}_t^{\mathcal{L}^{(j\independent i)}}$ that local DAG $\mathcal{G}^{\mathcal{L}^{(j\independent i)}}$ contains no edge from $V_t^i$ to $s_{t+1}^j$. It implies that the local DAG $\mathcal{G}^{\mathcal{L}^{(j\independent i)}}$ is strictly sparser than global DAG $\mathcal{G}$. With such property, we now can uses the local causal model to conduct the data augmentation to relief the sparsity problem in DRL RS.

Consider a counterfactual question \textit{``What if user $u$ interested in item $j$ instead of item $i$ at timestamp $t$?''}. It can be described in causal form - \textit{``What if component $s_t^i$ had value $x$ instead of $y$ at timestamp $t$?''}. It can be solved by using Pearl’s do-calculus to the causal model $\mathcal{M}$ to obtain a sub-model,
\begin{align}
    \mathcal{M}_{\text{do}(s_t^i=x)}^\mathcal{L} = (V,U,F_x) \text{ where } F_x = F \setminus f^i \cup\{s_t^i = x\}.
\end{align}
Moreover, the incoming edge to $s_t^i$ will be removed from $\mathcal{G}_{\text{do}(s_t^i=x)}$.
Now we will utilize the local causal model to generate data that unseen by agent but causally valid. In order to achieve that, we will augment the data based on the counterfactual modification with subset of the causal factor at timestamp $t$, and keep the reaming factors unchanged. Such augmentation process can use the counterfactual model $\mathcal{M}_{\text{do}(s_t^i=x)}^\mathcal{L}$ to modify the causal factors $s_t^{i\cdots j}$ and re-generate the corresponding children in the DAG which would increase the 

However, such process is computational expensive in our recommendation scenario as it requires re-sampling for the children in DAG. Inspired by the idea for collaborative filtering and the state composition form which was mentioned previously. We can simplify the process by omitting the sampling process. Specifically, the core of the augmentation is to estimate the $\mathcal{M}_{\text{do}(s_t^i=x)}^\mathcal{L}$. It can be easily to get by assuming that similar users will have similar interests which is the idea of collaborative filtering. Under such assumption, we can get $\mathcal{M}_{\text{do}(s_t^i=x)}^\mathcal{L}$ by replacing the causal independent component of $s_t$ using the local causal model. For example, those interaction histories that not affect current recommendation which could be recognized as causally independent with the current action $a_t$. 

The overall algorithm\footnote{For set-based representations} can be found in \Cref{alg:augmentation}.
\begin{algorithm}[ht]
\SetAlgoLined
 \SetKwInOut{Input}{input}
 \SetKwProg{Fn}{function}{ is}{end}
 \Input{Two trajectories $\tau_1$ and $\tau_2$ from replay buffer $D$}
 \Fn{Augmentation ($\tau_1,\tau_2$)}{
     $s_1,a_1,s_1' \leftarrow \tau_1$, $s_2,a_2,s_2'\leftarrow \tau_2$\;
     $m_1, m_2 \leftarrow$ ind($s_1, a_1$), ind($s_2, a_2$)\; 
     $d \leftarrow$ random sample from ($m_1 \cap m_2$)\;
     $s_g,a_g,s_g' = s_1,a_1,s_1'.$copy()\;
     $s_g[d],a_g[d],s_g'[d] = s_2[d],a_2[d],s_g'[d]$\;
     $\tau_g \leftarrow s_g,a_g,s_g',R(s_g,a_g)$\;
     $D.append(\tau_g)$ \textbf{if} $s_g[d], a_g[d], s'_g[d]$ exists \textbf{else} $D$\;
     \Return $D$;
 }
 \Fn{ind(s,a)}{
    Generate the adjacency matrix $M$ by using $s$ and $a$\;
    Find the connected (dependent) components set $C$\;
    \Return index of independent components in $\mathcal{G}\setminus  C$ \; 
 }
 \caption{Counterfactual Data Augmentation}
 \label{alg:augmentation}
\end{algorithm}

\subsection{Intrinsically Motivated Exploration}
\label{sec:kl}
The second aspect we use to address the sparsity is the intrinsically motivated exploration strategy. We propose to use the empowerment to represent the intrinsical motivation. It can boost agent's exploration capability so that more states can be reachable which can be used to produce corresponding potential informative interaction trajectories.
Empowerment is an information-theoretic method where an agent executes a sequence of $k$ actions $\textbf{a}^k \in \mathcal{A}$ when in state $s \in \mathcal{S}$ according to a explore policy $\pi_{\textit{empower}}(s,\textbf{a}^k)$ (we use $\pi_e(s,\textbf{a}^k)$ to shorten the notation) which is a conditional probability distribution: $\pi_e: \mathcal{S}\times\mathcal{A}\rightarrow [0,1]$. The agent’s aim is to identify an optimal policy $\pi_e$ that maximizes the mutual information $I[\textbf{a}^k, s'|s]$ between the action sequence $\textbf{a}^k$ and the state $s'$
to which the environment transitions after executing the sequence $\textbf{a}$ in current state $s$, formulated as:
\begin{align}
    \overline{\mathbb{E}}(s) & = \max_{\pi_e} I[\textbf{a}^k, s'|s] \\ &=  \max_{\pi_e}\mathbb{E}_{\pi_e(s,\textbf{a}^k)\mathcal{P}(s,\textbf{a}^k,s')}\log\bigg[\frac{p(\textbf{a}^k,s'|s)}{\pi_e(\textbf{a}^k,s)}\bigg].
    \label{eq1}
\end{align}
Here, $\overline{\mathbb{E}}(s)$ refers to the optimal empowerment value and $\mathcal{P}(s,\textbf{a}^k,s')$  to the probability of transitioning to $s'$ after  executing the action sequence $\textbf{a}^k$ in state $s$, where $\mathcal{P}:\mathcal{S}\times\mathcal{A}\times\mathcal{S} \rightarrow [0,1]$. 
Importantly, 
\begin{align}
    p(\textbf{a}^k,s'|s) = \frac{\mathcal{P}(s,\textbf{a}^k,s')\pi_e(\textbf{a}^k,s)}{\sum_{\textbf{a}^k} \mathcal{P}(s,\textbf{a}^k,s')\pi_e(\textbf{a}^k,s)}
\end{align}
is the inverse dynamics model of $\pi_e$. The optimal empowerment values are obtained by the policy $\pi^*$ that maximizes $\mathbb{E}^{\pi^*}(s)$.

However, the above definition of empowerment is more general than RL setting as it considers $k$-step policy but RL normally considers single step policy. Moreover, estimate the $k$-step empowerment is challenging. Hence, we use $k=1$ in this study to narrow down the empowerment into RL setting which only consider one step further. Blahut-Arimoto algorithm~\cite{arimoto1972algorithm,blahut1972computation} shows that empowerment can be solved in low-dimensional discrete settings.
Moreover,~\cite{mohamed2015variational} uses the parametric function approximators to estimate the empowerment in high-dimensional and continuous state-action spaces. It provides the theoretical guarantee for using empowerment in recommender systerms as the state-action spaces are high-dimensional~\cite{chen2021survey}. There are two possibilities for utilizing empowerment in RL:
\begin{itemize}
    \item Find high mutual information between actions and the subsequent state that achieved by that action.
    \item Train a behavioral policy to take an action in each state such that the expected empowerment value of the next state is highest.
\end{itemize}
Both approaches are feasible for normal reinforcement learning setting which can encourage agent to take an action that can result in the maximum number of future states. But there is some conceptual difference between them. Second approach seeks states with a large number of reachable next states~\cite{jung2011empowerment, leibfried2019unified}. The first approach aims to find high mutual information between actions and subsequent state which is not necessarily the same as seeking high empowered state~\cite{mohamed2015variational}. The first approach can be achieved by transferring state and its subsequent states' representations into KL divergence and minimizing it~\cite{kumar2018empowerment}. However, the transformation introduces extra complexity and information lost which may affect the performance. The second approach uses the behavior policy to explore the highly empowerment states would be more suitable and simple for our setting. The majority reason is that we are using model-free approach to solve the problem. The model-free RL methods maintain two policy which are target policy $\pi$ and behavior policy $\pi_e$. Second approach is more suitable for the model-free approaches as it does not require extra computational cost to traverse all of the subsequent states and calculate the KL divergence. It would be easily adopts into the existing RL frameworks.

Hence, the goal of the MDP process with the empowerment can be rewritten as,
\begin{align}
    \max_{\pi_b} \mathbb{E}_{\pi_b,\mathcal{P}} \bigg[\sum_{t=0}^\infty \gamma^t (\alpha\cdot R(s_t,a_t) + \beta \cdot \frac{p(a_t|s_{t+1},s_t)}{\pi_b(a_t,s_t)}\bigg]
    \label{eq2}
\end{align}
where $\pi_b$ is the behavior policy, $\alpha, \beta$ are constants used to balance instant reward and empowerment. We use the empowerment as the extra term to the reward signal $R(s_t,a_t)$ to encourage agent for exploration.
\subsection{Training Procedure}
From information-theoretic prospective,  optimizing for empowerment is equivalent to optimize the inverse dynamics~\cite{leibfried2019unified,cover1999elements} based on the distribution $\pi_e(s,a)$. Hence, we introduce the inverse dynamics into the objective function to calculate the empowerment. Our method is build up on Soft Actor-Critic (SAC)~\cite{haarnoja2018soft} with temperature tuning and deterministic policy. The overall training algorithm can be found in ~\Cref{alg:train}.

We follow the same training strategy as standard SAC algorithm. However, as the empowerment is introduced, we modify the objective function to ensure the empowerment term can be optimized. We use several function approximators to learn different components in the proposed method. Value function $V$ is parametrized by $\psi$, Q-function is parametrized by $\theta$, target policy is parametrized by $\phi$ and inverse dynamics is parameterized by $\xi$. As we are using an off-policy algorithm where the transition probability is not be learned, we use the $\mathcal{P}$ to represent the state transition probability in the environment. The soft Q-function can be trained by minimizing the following objective function:
\begin{align}
    J_Q(\theta) = \mathbb{E}_{(s_t,a_t)\sim\mathcal{D}}\big[Q_\theta(s_t,a_t) - (r(s_t,a_t) + \gamma V_{\psi}(s_{t+1}))^2\big].
\end{align}
The target function $V_\psi$ can be optimized by minimizing:
\begin{align}
    J_V(\psi) = \mathbb{E}_{s_t\sim \mathcal{D}} \Big[V_\psi(s_t) -\mathbb{E}_{a_t\sim\pi_\phi}\big(Q_\theta(s_t,a_t)+\underbrace{\beta g(s_t,a_t)}_{\text{policy}}\big)^2\Big],
\end{align}
where $\beta$ is a constant used to balance the empowerment.
Different from the origin SAC algorithm, we replace the policy term from $-\log\pi_\phi(s_t,a_t)$ to $g(s_t,a_t)$ to consider the empowerment where $g(s_t,a_t)$ is defined as:
\begin{align}
    g(s_t,a_t) = \mathbb{E}_{\mathcal{P}(s'|s_t,a_t)}\big[\log p_\xi(a_t|s',s_t) -\log \pi_\phi(s_t,a_t)\big].
\end{align}
Note that, different from $s_t$, the $s'$ represent all the possible subsequent states where $a_t$ is executed in state $s_t$ at timestamp $t$.
\begin{algorithm}[ht]
 \SetAlgoLined
 \SetNoFillComment
 \newcommand\mycommfont[1]{\footnotesize\ttfamily\textcolor{blue}{#1}}
 \SetCommentSty{mycommfont}
 \SetKwInOut{Input}{input}
 \Input{Parameters $\phi$ for policy, $\theta$ for Q-networks, $\psi$ for value function, $\xi$ for inverse dynamic, replay buffer $\mathcal{D}$,  threshold $T$.}
 \For{each episode till converge}{
    $s_0 \leftarrow$ initiate state \; 
    \For{each environment step}{
        $a_t \sim \pi_{\phi}(a_t|s_t)$\;
        $r_t \leftarrow \mathcal{R}(s_t,a_t)$\;
        $s_{t+1} \leftarrow \mathcal{P}(s_{t+1}|s_t,a_t)$\tcc*{Sample from environment}
        $\mathcal{D} \leftarrow \mathcal{D} \cup \{s_t,a_t,r_t,s_{t+1}\}$\;
        \If{$r_t \geq T$}{
        Augmentation($\mathcal{D}$) \;
        }
        Update $T$ based on \Cref{eq:threhold}\;
        
    }
    \For{each gradient step}{
    $\theta \leftarrow \theta - \lambda_Q \nabla_{\theta} J_Q(\theta)$ \; 
    $\psi \leftarrow \psi - \lambda_\psi \nabla_\psi J_V(\psi)$\;  
    $\phi \leftarrow \phi - \lambda_\phi \nabla_\phi J_\pi(\phi)$\; 
    $\xi \leftarrow \xi - \lambda_\xi \nabla_\xi J_p(\xi)$\; 
    $\alpha \leftarrow \alpha - \lambda\nabla_\alpha J(\alpha)$ \; 
    }
 }
 \caption{Overall training algorithm}
 \label{alg:train}
\end{algorithm}

Similarly, the optimization of the policy $\pi(\phi)$ can be written as:
\begin{align}
   J_\pi(\phi) = -\mathbb{E}_{s_t\sim \mathcal{D}}\Big[\mathbb{E}_{a_t\sim\pi_\phi}\big[\beta g(s_t,a_t)+Q_\theta(s_t,a_t)\big]\Big],
\end{align}
where apply the same substitution. The inverse dynamic $p(\xi)$ will be updated based on:
\begin{align}
    J_p(\xi) = -\mathbb{E}_{\pi_\phi}\big[\log{p_\xi}(a_t|s',s_t)\big].
\end{align}
Lastly, the temperature parameter will be adjusted automatically by using the following entropy method~\cite{haarnoja2018softa}:
\begin{align}
    J(\alpha) = -\mathbb{E}_{a_t\sim\pi_\phi}\big[\alpha \log\pi_\phi(s_t,a_t) + \alpha \mathcal{H}\big].
\end{align}
Note that, SAC uses exponentially averaged value $\psi'$ to stabilize the training process~\cite{mnih2015human}. The update rule can be written as: $\psi' \leftarrow \lambda_{\psi'} \psi+ (1-\lambda_{\psi'})\psi'$.

Moreover, we conduct the data augmentation to the replay buffer after each interaction to generate causally valid unseen trajectories. Such augmentation can provide more trajectories at the early stage to increase the samples. Specifically, most of model parameters are learned by sampling from the replay buffer $\mathcal{D}$. The training process can be described as searching states or state-action pairs in the $\mathcal{D}$ to update the target policy such that the received reward is maximized. As the augmentation introduces more samples into the replay buffer, the gradient update process have a higher chance to achieve a better policy. 

It is worth to mention that, we are only augmenting those informative trajectories. However, the definition of the informative trajectories are highly depends on the learning progress.
We believe that every trajectories with non-zero reward are informative at the early stage but are harmful when the final target policy close optimal. Hence, we selectively conduct the augmentation with replay buffer to ensure that those zero-reward trajectories will not be augmented to increase sparse. However, as the interaction goes forward, the way we determine informative trajectories is changing. Some trajectories is informative at the early stage as agent need to explore all the possibilities. At the later stage, agent will pursue higher rewarding trajectories which makes those low-rewarding trajectories less useful. In such situation, we design a adaptive threshold to evaluate the trajectory is worth for augmentation or not. The designed adaptive threshold is intuitive where moving average is used. It can be represented as,
\begin{align}
    T = \sigma/\lambda_d \label{eq:threhold} \text{ if } T \leq T_{max} \text{ else } T_{max}
\end{align}
where $\sigma$ is a customized constant to determinate the initial value of the threshold and the decay rate $\lambda_d \in (0,1]$. The decay rate will decrease when number of episode increase. With a set of value $(\sigma, \lambda_d)$, we can achieve a monotone increasing threshold. Ideally, we start with the $\sigma = 1$ and $\lambda_d = 1$ as initial values. $T_{max}$ is an environment constant which represents the maximum reward that agent can achieve each step.
\section{Experiments}
In this section, we conduct the experiments to answer three main research questions:
\begin{itemize}
    \item \textbf{RQ1}: Does IMRL outperform than existing DRL approaches in both offline and online settings?
    \item \textbf{RQ2}: Can IMRL help to relief the sparsity interaction problem in DRL RS in online simulation environments?
    \item \textbf{RQ3}: How each components contribute to the final performance in online simulation environments?
\end{itemize}
\subsection{Experiment Setup}
In order to demonstrate the superiority of IMRL, we've conducted the experiments on both offline setting and online simulation setting.

\subsubsection{Offline Datasets} We use six public available datasets: 
\begin{itemize}
    \item \textbf{MoveLens-20M}\footnote{https://grouplens.org/datasets/movielens/20m/} is a dataset about the user behavior of watching movies.
    \item \textbf{Librarything}\footnote{\url{http://cseweb.ucsd.edu/~jmcauley/datasets.html}\#social\_data} is a dataset about book review information.
    \item \textbf{Book-crossing}\footnote{http://www2.informatik.uni-freiburg.de/~cziegler/BX/}is a dataset related to book preference.
    \item \textbf{Netflix Prize\footnote{\url{https://www.kaggle.com/netflix-inc/netflix-prize-data}}} is a dataset from Netflix yearly competition for recommendation. 
    \item \textbf{Amazon-CD}\footnote{https://jmcauley.ucsd.edu/data/amazon/} is e-commerce datasets which contains user's purchase behavior.
    \item \textbf{GoodReads}\footnote{https://www.goodreads.com/} is a book dataset.
\end{itemize}
The statistics of those datasets are summarized in~\Cref{tab:stat}.
\begin{table}[ht]
    \centering
    \caption{Statistics of the datasets used in our offline experiments.}
    \begin{tabular}{c|c|c|c|c}
    \hline 
        Dataset & \# of Users & \# of Items & \# of Inter. & Density  \\\hline
         Amazon CD & 75,258 & 64,443 & 3,749,004 & 0.08\% \\
         Librarything & 73,882 & 337,561 & 979,053 & 0.004\% \\
         Book-Crossing & 278,858 & 271,379  & 1,149,780 & 0.0041\%\\
         GoodReads & 808,749 &  1,561,465 & 225,394,930 & 0.02\%\\
         MovieLens-20M &  138,493  & 27,278 & 20,000,263 & 0.53\%\\ 
         Netflix & 480,189  & 17,770 & 100,498,277 & 1.18\%\\
    \hline 
    \end{tabular}
    \label{tab:stat}
\end{table}
Moreover, due to the special interaction logic in reinforcement learning based methods. Extra data preparation process is required to ensure that the agent can interact with such offline datasets. We follow the same strategy in previous work~\cite{chen2019generative} that transfer those datasets into reinforcement learning environments so that IMRL can interact with.
\subsubsection{Baselines and offline evaluation metrics}
The following baselines are selected which contain both non-reinforcement learning based methods and reinforcement learning based methods:
\begin{itemize}
    \item SASRec~\cite{kang2018self} is a well-known baseline for sequential recommendation method that utilize the self-attention mechanism.
    \item CASR~\cite{wang2021counterfactual} is a counterfactual data augmentation method for sequential recommendation. As CASR only conduct the augmentation, we select the STAMP~\cite{zhao2018recommendations} to make recommedation which is described in CASR. 
    \item CauseRec~\cite{zhang2021causerec} is a counterfactual sequence generation method for sequential recommendation.
    \item CoCoRec~\cite{cai2021category} is a category-aware collaborative method for sequential recommendation.
    \item DEERS~\cite{zhao2018recommendations} is a reinforcement learning based recommendation method that considers both positive and negative feedback.
    \item KGRL~\cite{chen2020knowledge} is a reinforcement learning based method that utilize the capability of GCN to process the knowledge graph information.
    \item TPGR~\cite{chen2019large} is a model that uses reinforcement learning and binary tree for large-scale interactive recommendation.
    \item PGPR~\cite{xian2019reinforcement}is a knowledge-aware model that employs reinforcement learning for explainable recommendation.
\end{itemize}
It is worth to mention that, because of the different training paradigm of those two kinds of methods (i.e., supervised learning and reinforcement learning), we are not able to guarantee that the comparison with those existing non-reinforcement learning based state-of-the-art methods are strictly fair. We've conducted those supervised learning based methods in the same setting as well as those reinforcement learning based methods.
Recall, precision, nDCG are selected as the evaluation metrics.
\begin{table*}[!ht]
\caption{The overall results of our model comparison with several state-of-the-art models in different datasets. The result was reported based on top-20 recommendation and highest results are in bold and second highest is marked by *}\smallskip
\begin{minipage}[ht]{1.0\linewidth}
\resizebox{\columnwidth}{!}{%
\begin{tabular}{ccccccc}
\hline
\multicolumn{1}{c|}{Dataset} & \multicolumn{3}{c|}{Amazon CD} & \multicolumn{3}{c}{Librarything} \\ \hline
\multicolumn{1}{c|}{Measure (\%)} & \multicolumn{1}{c|}{Recall} & \multicolumn{1}{c|}{Precision} & \multicolumn{1}{c|}{nDCG} & \multicolumn{1}{c|}{Recall} & \multicolumn{1}{c|}{Precision} & nDCG \\ \hline
\multicolumn{1}{c|}{SASRec} & 5.129 $\pm$ 0.233 & 2.349 $\pm$ 0.144 & \multicolumn{1}{c|}{4.591 $\pm$ 0.312 } & 8.419 $\pm$ 0.294 & 6.726 $\pm$ 0.139  & 7.471 $\pm$ 0.221  \\ 
\multicolumn{1}{c|}{CASR} & 8.321 $\pm$ 0.212 & 5.012 $\pm$ 0.129* &  \multicolumn{1}{c|}{7.282 $\pm$ 0.212*} & 14.213 $\pm$ 0.311 & 12.512 $\pm$ 0.219 & 13.555 $\pm$ 0.198  \\ 
\multicolumn{1}{c|}{CauseRec} & 9.124 $\pm$ 0.213* & 4.892 $\pm$ 0.299 & \multicolumn{1}{c|}{6.214 $\pm$ 0.479} & 14.222 $\pm$ 0.421  & 12.582 $\pm$ 0.321*  & 12.875 $\pm$ 0.317    \\  
\multicolumn{1}{c|}{CoCoRec} & 8.982 $\pm$ 0.221 & 4.982 $\pm$ 0.312 & \multicolumn{1}{c|}{7.122 $\pm$ 0.218} & 13.982 $\pm$ 0.123 &  11.233 $\pm$ 0.300 & 12.098 $\pm$ 0.302   \\  
\multicolumn{1}{c|}{DEERS} & 7.123$\pm$ 0.221 & 3.581 $\pm$ 0.200 & \multicolumn{1}{c|}{6.341 $\pm$ 0.312} & 10.422 $\pm$ 0.231 & 10.321 $\pm$ 0.355  & 11.872 $\pm$ 0.241 \\  
\multicolumn{1}{c|}{KGRL} & 8.208 $\pm$ 0.241 & 4.782 $\pm$ 0.341 & \multicolumn{1}{c|}{6.876 $\pm$ 0.511} & 12.128 $\pm$ 0.241 & 12.451 $\pm$ 0.242 & 13.925 $\pm$ 0.252* \\  
\multicolumn{1}{c|}{TPGR} & 7.294 $\pm$ 0.312 & 2.872 $\pm$ 0.531 & \multicolumn{1}{c|}{6.128 $\pm$ 0.541} & 14.713 $\pm$ 0.644* & 12.410 $\pm$ 0.612 & 13.225 $\pm$ 0.722 \\  
\multicolumn{1}{c|}{PGPR} & 6.619 $\pm$ 0.123 & 1.892 $\pm$ 0.143 & \multicolumn{1}{c|}{5.970 $\pm$ 0.131 } & 11.531 $\pm$ 0.241 & 10.333 $\pm$ 0.341 & 12.641 $\pm$ 0.442  \\  
\hline
\multicolumn{1}{c|}{Ours} & \textbf{9.213 $\pm$ 0.219} & \textbf{5.032 $\pm$ 0.125} & \multicolumn{1}{c|}{\textbf{7.421 $\pm$ 0.231}} & \textbf{14.829 $\pm$ 0.321} & \textbf{12.610 $\pm$ 0.231} & \textbf{14.021 $\pm$ 0.335} \\ 
\hline
\end{tabular}%
}
\end{minipage}

\begin{minipage}[ht]{1.0\linewidth}
\resizebox{\columnwidth}{!}{%
\begin{tabular}{ccccccc}
\hline
\multicolumn{1}{c|}{Dataset} &  \multicolumn{3}{c|}{Book-Crossing} & \multicolumn{3}{c}{GoodReads} \\ \hline
\multicolumn{1}{c|}{Measure (\%)} & \multicolumn{1}{c|}{Recall} & \multicolumn{1}{c|}{Precision} & \multicolumn{1}{c|}{nDCG} & \multicolumn{1}{c|}{Recall} & \multicolumn{1}{c|}{Precision} & nDCG \\ \hline
\multicolumn{1}{c|}{SASRec}& 5.831 $\pm$ 0.272  &  3.184 $\pm$ 0.149 &\multicolumn{1}{c|}{4.129 $\pm$ 0.390} & 6.921 $\pm$ 0.312 & 5.242 $\pm$ 0.211 & 6.124 $\pm$ 0.210  \\ 
\multicolumn{1}{c|}{CASR}& 8.322 $\pm$ 0.300 & 5.012 $\pm$ 0.211  & \multicolumn{1}{c|}{5.922 $\pm$ 0.198 } & 11.228 $\pm$ 0.123 & 10.922 $\pm$ 0.339 & 10.233 $\pm$ 0.210   \\ 
\multicolumn{1}{c|}{CauseRec} &  9.213 $\pm$ 0.213 & \textbf{6.213 $\pm$ 0.198} &  \multicolumn{1}{c|}{6.872 $\pm$ 0.212} & 11.827 $\pm$ 0.431* & 10.982 $\pm$ 0.412  & 10.277 $\pm$ 0.312 \\  
\multicolumn{1}{c|}{CoCoRec} & 8.234 $\pm$ 0.231 & 5.182 $\pm$ 0.200 & \multicolumn{1}{c|}{5.829 $\pm$ 0.120} & 10.882 $\pm$ 0.233  & 10.012 $\pm$ 0.210 & 10.012 $\pm$ 0.129  \\  
\multicolumn{1}{c|}{DEERS} & 7.321 $\pm$ 0.320 & 2.574 $\pm$ 0.201  & \multicolumn{1}{c|}{6.123 $\pm$ 0.123} & 8.231 $\pm$ 0.122  & 9.318 $\pm$ 0.132  & 9.401 $\pm$ 0.184  \\  
\multicolumn{1}{c|}{KGRL} & 8.004 $\pm$ 0.223 & 3.521 $\pm$ 0.332 & \multicolumn{1}{c|}{7.641 $\pm$ 0.446} & 7.459 $\pm$ 0.401 & 11.444 $\pm$ 0.321*& 10.331 $\pm$ 0.331*  \\  
\multicolumn{1}{c|}{TPGR} & 7.246 $\pm$ 0.321 & 4.523 $\pm$ 0.442 & \multicolumn{1}{c|}{7.870 $\pm$ 0.412*} & 11.219 $\pm$ 0.323 & 10.322 $\pm$ 0.442 & 9.825 $\pm$ 0.642 \\  
\multicolumn{1}{c|}{PGPR} & 6.998 $\pm$ 0.112 & 3.932 $\pm$ 0.121 & \multicolumn{1}{c|}{7.333 $\pm$ 0.133} & 11.421 $\pm$ 0.223 & 10.042 $\pm$ 0.212 & 9.234 $\pm$ 0.242 \\  
\hline
\multicolumn{1}{c|}{Ours} & \textbf{9.331 $\pm$ 0.213} &  5.442 $\pm$ 0.124*& \multicolumn{1}{c|}{\textbf{7.921 $\pm$ 0.200}} & \textbf{12.013 $\pm$ 0.201} & \textbf{11.726 $\pm$ 0.138} & \textbf{10.612 $\pm$ 0.320}\\ 
\hline
\end{tabular}%
}
\end{minipage}

\begin{minipage}[ht]{1.0\linewidth}
\resizebox{\columnwidth}{!}{%
\begin{tabular}{ccccccc}
\hline
\multicolumn{1}{c|}{Dataset} &  \multicolumn{3}{c|}{MovieLens-20M} & \multicolumn{3}{c}{Netflix} \\ \hline
\multicolumn{1}{c|}{Measure (\%)} & \multicolumn{1}{c|}{Recall} & \multicolumn{1}{c|}{Precision} & \multicolumn{1}{c|}{nDCG} & \multicolumn{1}{c|}{Recall} & \multicolumn{1}{c|}{Precision} & nDCG \\ \hline
\multicolumn{1}{c|}{SASRec} & 14.512 $\pm$ 0.510& 12.412 $\pm$ 0.333& \multicolumn{1}{c|}{12.401 $\pm$ 0.422} & 11.321 $\pm$ 0.231 & 10.322 $\pm$ 0.294   & 14.225 $\pm$ 0.421 \\ 
\multicolumn{1}{c|}{CASR} & 17.324 $\pm$ 0.212 & 14.021 $\pm$ 0.210  & \multicolumn{1}{c|}{14.821 $\pm$ 0.213*} & 13.551 $\pm$ 0.240 & 12.412 $\pm$ 0.122  &  15.212 $\pm$ 0.211\\ 
\multicolumn{1}{c|}{CauseRec}  & 17.625 $\pm$ 0.331*  &  14.982 $\pm$ 0.291* & \multicolumn{1}{c|}{14.231 $\pm$ 0.211} & 13.982 $\pm$ 0.325* & 12.842 $\pm$ 0.222* & \textbf{15.882 $\pm$ 0.261} \\  
\multicolumn{1}{c|}{CoCoRec}  & 16.212 $\pm$ 0.211 & 14.222 $\pm$ 0.290 & \multicolumn{1}{c|}{13.491 $\pm$ 0.219} & 13.762 $\pm$ 0.199 & 12.001 $\pm$ 0.129  & 13.284 $\pm$ 0.235  \\  
\multicolumn{1}{c|}{DEERS} & 16.123 $\pm$ 0.312 & 12.984 $\pm$ 0.221 & \multicolumn{1}{c|}{12.322 $\pm$ 0.198 } & 12.847$\pm$ 0.219 & 11.321 $\pm$ 0.294 & 14.521 $\pm$ 0.401 \\  
\multicolumn{1}{c|}{KGRL}  & 16.021 $\pm$ 0.498  & 14.989 $\pm$ 0.432 & \multicolumn{1}{c|}{13.007 $\pm$ 0.543} & 13.009 $\pm$ 0.343 & 11.874 $\pm$ 0.232 &13.082 $\pm$ 0.348 \\  
\multicolumn{1}{c|}{TPGR}  & 16.431 $\pm$ 0.369 & 13.421 $\pm$ 0.257 & \multicolumn{1}{c|}{13.512 $\pm$ 0.484} & 12.512 $\pm$ 0.556 & 11.512 $\pm$ 0.595 & 15.425 $\pm$ 0.602 \\  
\multicolumn{1}{c|}{PGPR} & 14.234 $\pm$ 0.207 & 9.531 $\pm$ 0.219 & \multicolumn{1}{c|}{11.561 $\pm$ 0.228} & 10.982 $\pm$ 0.181 & 10.123 $\pm$ 0.227 & 15.134 $\pm$ 0.243 \\  
\hline
\multicolumn{1}{c|}{Ours} & \textbf{17.798 $\pm$ 0.231} & \textbf{15.041 $\pm$ 0.122} & \multicolumn{1}{c|}{\textbf{14.991 $\pm$ 0.132 }} & \textbf{14.421 $\pm$ 0.239} & \textbf{13.012 $\pm$ 0.321} & 15.448 $\pm$ 0.122* \\ 
\hline
\end{tabular}%
}
\end{minipage}
\label{tab:result}
\end{table*}
\subsubsection{Online Simulation}
Different from offline datasets, online simulation platforms are all based on gym\footnote{https://gym.openai.com} which is a standard toolkit for reinforcement learning research.
We conduct the online experiments on three widely used public simulation platforms: VirtualTB~\cite{shi2019virtual}, RecSim~\cite{ie2019recsim} and RecoGym~\cite{rohde2018recogym}, which mimic online recommendations in real-world applications.

\vspace{2mm}
\noindent{\bf VirtualTB} is a real-time simulation platform for recommendation, where the agent recommend items based on users' dynamic interests. VirtualTB uses a pre-trained generative adversarial imitation learning (GAIL) to generate different users who have both static interest and dynamic interest. Moreover, the interactions between users and items are generated by GAIL as well. Benefit from that, VirualTB can provide a large number of users and the corresponding interactions to simulate the real-world scenario. VirtualTB would generate different users each time after the initialization and the dynamic interest will change after each single interaction.

\vspace{2mm}
\noindent{\bf RecSim} is a configurable platform for authoring simulation environments that naturally supports sequential interaction with users in recommender systems. RecSim differs from VirtualTB in containing different, simpler tasks but fewer users and items. The task we decide to use in RecSim, namely interest evolution. The interest evolution encourages the agent to explore and fulfill the user's interest without further exploitation. 
\begin{figure*}[!ht]
    \centering
    \begin{subfigure}{0.32\linewidth}
        \includegraphics[width=\linewidth]{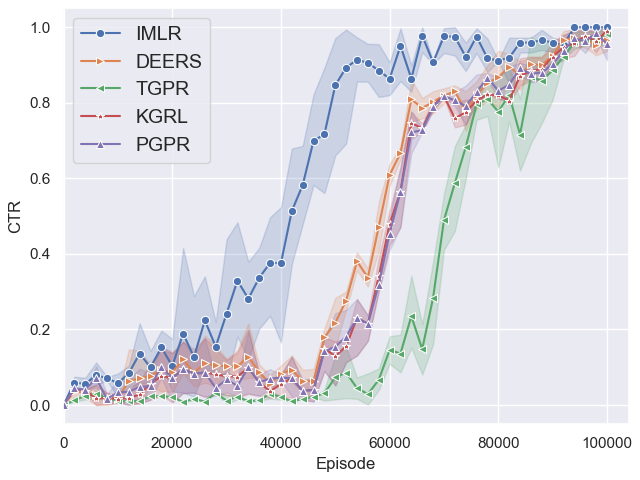}
        \caption{VirtualTB}
        \label{fig:virtualtb}
    \end{subfigure}
    \begin{subfigure}{0.32\linewidth}
        \includegraphics[width=\linewidth]{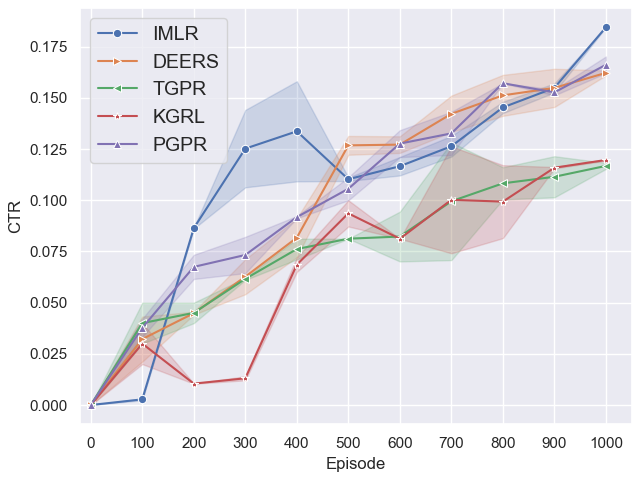}
        \caption{RecoGym}
        \label{fig:RecoGym}
    \end{subfigure} 
    \begin{subfigure}{0.32\linewidth}
        \includegraphics[width=\linewidth]{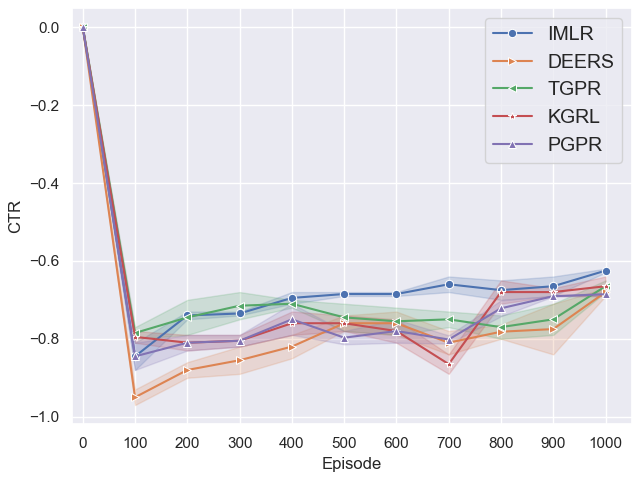}
        \caption{RecSim}
        \label{fig:recsim}
    \end{subfigure}
   
    \caption{Overall results for online simulation environments. Episode represents the number of test episode.}
    \label{fig:result_online}
\end{figure*}

\vspace{2mm}
\noindent{\bf RecoGym} is a small platform, where users have no long-term goals. Different from RecSim and VirtualTB, RecoGym is designed for computational advertising. Similar with RecSim, RecoGym uses the click or not to represent the reward signal. Moreover, similar with RecSim, users in those two environments do not contain any dynamic interests.

\subsubsection{Baselines for online simulation}
In our online simulation experiments, all the baselines are reinforcement learning based. Hence, those non-reinforcement learning based methods are ignored as they are not able to interact with the gym-based environment.
It is worth to mention that, some methods require extra side information from the environment which is not exist in those three platforms. Hence, we have to remove those components to ensure the comparison is fair (i.e., every method receive the same kind of state representation). The major evaluation metric used for online simulation is determined by the platform which is the Click-Through-Rate (CTR).

IMRL is implemented by using Pytorch~\cite{paszke2019pytorch} and all experiments are conducted on a server with two Intel Xeon E5-2697 v2 CPUs with 4 NVIDIA TITAN X Pascal GPUs, 2 NVIDIA TITAN RTX, 2 NVIDIA RTX A5000 and 768 GB memory.
We provide details about model parameters for reproducibility. We set the hidden unit to 256 for the actor network and the critic network, respectively. Learning rate, $\gamma$, and size of replay buffer are set to $0.0003$, $0.99$ and $1e^6$, respectively, during experiments. The training episode is set to 1e6 and test is conducted every 10 episodes in VirtualTB. And the training episode is set to $10,000$ for RecoGym and RecSim and tests are condcuted every 10 episodes.

\subsection{Offline Experiments}
The fully results could be found in~\Cref{tab:result}. We find that our method IMRL are generally outperforms than all those existing state-of-the-art methods both in non-reinforcement learning based methods and reinforcement learning methods. Notice that IMRL does not beat CauseRec in two datasets but still outperform than all the others. Although IMRL does not has a better precision than CauseRec in Book-Crossing, we can find that recall and nDCG is better than CauseRec. The same situation also happens in Netflix where the nDCG of IMRL is lower than CauseRec while precision and recall is better than CauseRec. 

\subsection{Online Simulations (RQ2)}
We also report the performance of those selected reinforcement learning based baselines in three online simulation environments. The results can be found in~\Cref{fig:result_online}. As we can find that, IMRL is outperform than all the others in all of selected three simulation platforms. The performance in RecoGym and RecSim are quite close as those two environment are very small where does not require a complex exploration policy. Hence, we will focus on the later discussion in VirtualTB as it has a more complex environment which is more similar with the real-world situation. 

The simplest way to evaluate the sparsity is be relived or not is to evaluate the speed of the model that tend to converge. In reinforcement learning, the way we use to measure the sparsity is  the number of useful samples are fed into the agent via replay buffer or sampled from the environment. Hence, dense environment can boost the model to converge at the early stage. In~\Cref{fig:virtualtb}, we can find that IMRL has a outstanding speed of convergence than other methods in VirtualTB which shows that it can overcome the sparse environment. In RecoGym and RecSim, IMRL also demonstrates a considerable improvement when comparing with those baselines. The majority reason is that, RecoGym and RecSim are small environment which contains only a few items and users. The sparsity is not serious and can be handled by random exploration.
\begin{figure}[!ht]
    \centering
    \includegraphics[width=\linewidth]{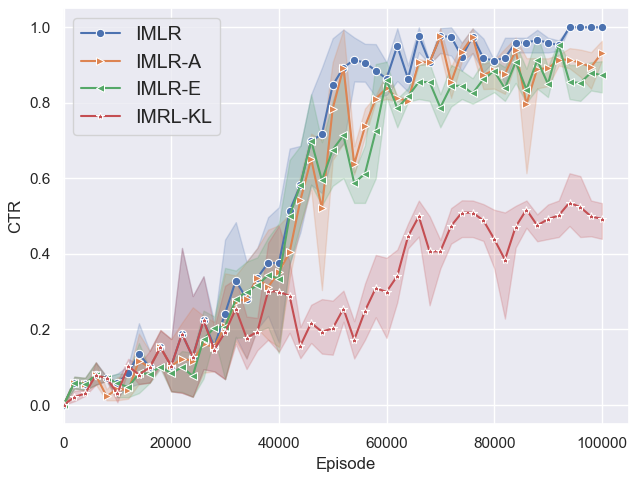}
    \caption{Ablation study on VirtualTB.}
    \label{fig:ablation}
\end{figure}
\subsection{Ablation Study (RQ3)}
In order to answer RQ3, we conduct the experiments with the two major components in IMRL which are empowerment and augmentation. The result of such study can be found in~\Cref{fig:ablation} where IMRL-E denotes IMRL without empowerment and IMRL-A denotes IMRL without augmentation. Moreover, we also investigate the effect of different strategy of empowerment in IMRL. Specifically, we also investigate the KL-divergence approach mentioned in Section~\Cref{sec:kl}. We use IMRL-KL to represent such method.
We can find that both components play an important role in IMRL and contribute to the final performance jointly. Furthermore, we can see that IMRL-KL does not perform as well as others. One of the possible reason is that the information lost during the transformation during the calculation of KL-divergence. Hence, we can infer that our approach of using the empowerment is better than KL-Divergence.
\section{Related Work}
In this section, we will briefly review two topics which related to our work: reinforcement learning based recommendation and causality in recommender systems.

\noindent\textbf{Reinforcement learning based recommendation}.~\cite{zheng2018drn} introduce the DRL into RS by using DQN to conduct the news recommendation. It uses double DQN to build the user's profile and design a activeness score to evaluate user is active or not.~\cite{zhao2018recommendations} extend this method by introducing the negative feedback.
~\cite{chen2019generative} uses cascading DQN and design generative user model method to handle the unknown reward situation.~\cite{chen2019top} design a scaleable policy-gradient based methods for recommendation by introducing a policy correction gradient estimator to reduce the variance.~\cite{xu2020reinforcement} design a Pairwise Policy Gradient method for recommendation to reduce the variance as well.~\cite{chen2019large} propose a tree-based method for large-scale interactive recommendation by utilizing the actor-critic algorithm.~\cite{chen2020knowledge} integrates the knowledge graph into the actor-critic structure and uses graph convolutional network to capture the information.~\cite{xian2019reinforcement} design a knowledge graph based environment for explainable recommendation.~\cite{chen2021generative} uses the inverse reinforcement learning to avoid the elaborate reward function in online recommendation.~\cite{he2020learning} uses SAC to conduct the multi-module recommendation via multi-agent approach.

\noindent\textbf{Causality in recommender systems}.  Causality receives significant research interest in recent literature about recommendation. It has been widely used for debias or data augmentation for RS.~\cite{wei2021model} use model-agnostic counterfactual reasoning to address the popularity bias in RS. Similarly,~\cite{zhang2021causal} propose causal intervention to relief the popularity bias as well.~\cite{wang2021counterfactual} design a counterfactual data augmentation methods by measuring embedding difference to generate new user sequence. Differently,~\cite{zhang2021causerec} separate the users' historical actions into dispensable and indispensable items where dispensable items can be omitted without affecting the final recommendation results. New user sequences are generated by replacing the dispensable items. In recent years, causality shows a strong connection with the reinforcement learning as both of them can affect the status of the input~\cite{gershman2017reinforcement}.~\cite{zhu2019causal} use the reinforcement learning to conduct the causal discovery where employs actor-critic algorithm to discover different DAG structure.~\cite{dasgupta2019causal} propose a meta reinforcement learning framework to conduct the causal reasoning by exploring different causal structures.~\cite{zhang2020causal} uses causal inference to determine the unobserved confounders to improve the performance of the imitation learning.~\cite{madumal2020explainable} utilizes casual inference to build an explainable reinforcement learning model.

\section{Conclusion}
In this paper, we propose IMRL to address the sparse interaction problem in DRL RS from two folds which are quantity and quality. We proposes a counterfactual based method to augment informative interaction trajectories and empowerment based exploration to boost the possibility of finding high quality trajectories. We have conducted the experiments on both offline datasets and online simulation platforms to demonstrate the superiority of the proposed method.
In the future, we are planning to further investigate power of empowerment and new solutions for reliving the sparse interaction for DRL RS. Moreover, how to stabilize the training process of DRL in RS is another challenges that we are targeting.
\bibliographystyle{IEEEtran}
\bibliography{IEEEabrv,sample}

\begin{IEEEbiography}[{\includegraphics[width=1in,height=1.25in,clip,keepaspectratio]{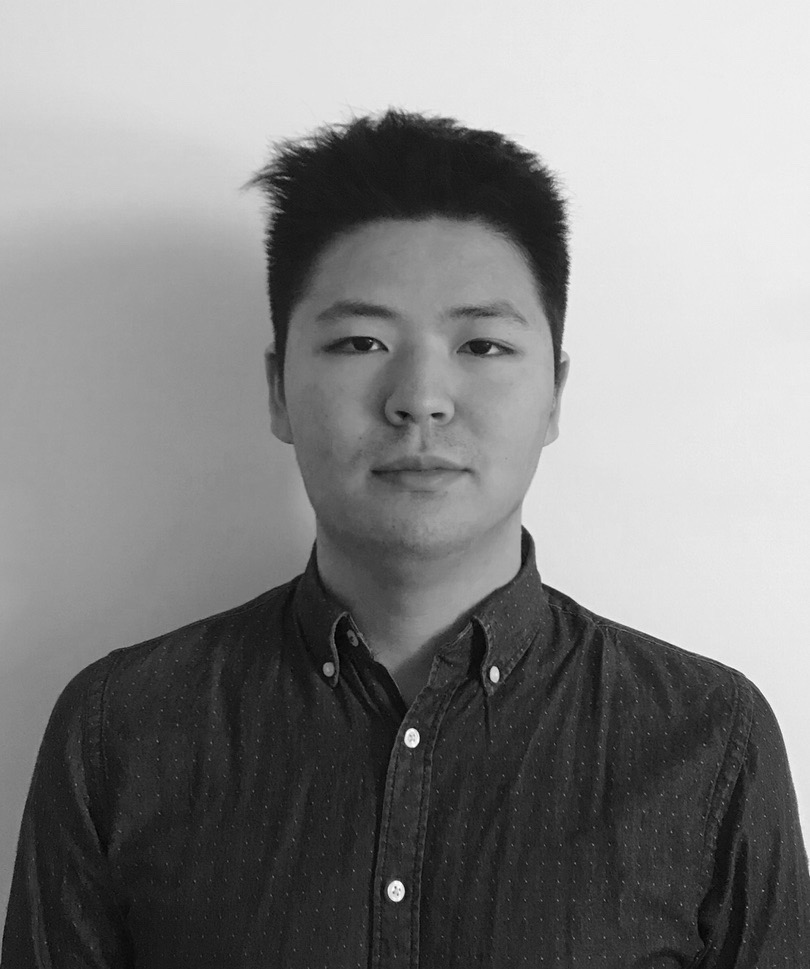}}]{Xiaocong Chen}
is a PhD student with School of Computer
Science and Engineering, The University of
New South Wales (UNSW), Australia. His research
interests are deep learning application and recommender systems.
\end{IEEEbiography}

\begin{IEEEbiographynophoto}{Siyu Wang}
is a PhD student with School of Computer
Science and Engineering, The University of
New South Wales (UNSW), Australia. Her research
interests are casual inference and recommender systems.
\end{IEEEbiographynophoto}

\begin{IEEEbiography}[{\includegraphics[width=1in,height=1.25in,clip,keepaspectratio]{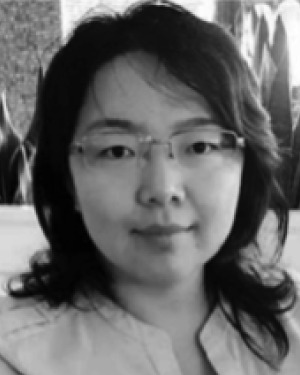}}]{Lina Yao} received the master’s and PhD degrees from School of Computer Science, University of Adelaide, in 2014. She is currently an Associate Professor with School of Computer Science and Engineering, The University of New South Wales (UNSW), Australia. Her research interest lies in machine learning and data mining, and applications to Internet of Things, information filtering and recommending, human activity recognition and brain computer interface. She is a senior member of the IEEE and the ACM. 
\end{IEEEbiography}

\begin{IEEEbiography}[{\includegraphics[width=1in,height=1.25in,clip,keepaspectratio]{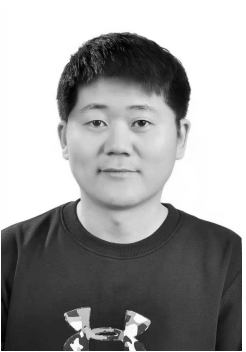}}]{Lianyong Qi} received his PhD degree in Department of Computer Science and Technology from
Nanjing University, China. In 2010, he visited the
Department of Information and Communication
Technology, Swinburne University of Technology, Australia. Now, he is a professor of the
College of Computer Science and Technology,
China University of Petroleum, China. His research interests include big data, recommender
systems and service computing.
\end{IEEEbiography}

\begin{IEEEbiography}[{\includegraphics[width=1in,height=1.25in,clip,keepaspectratio]{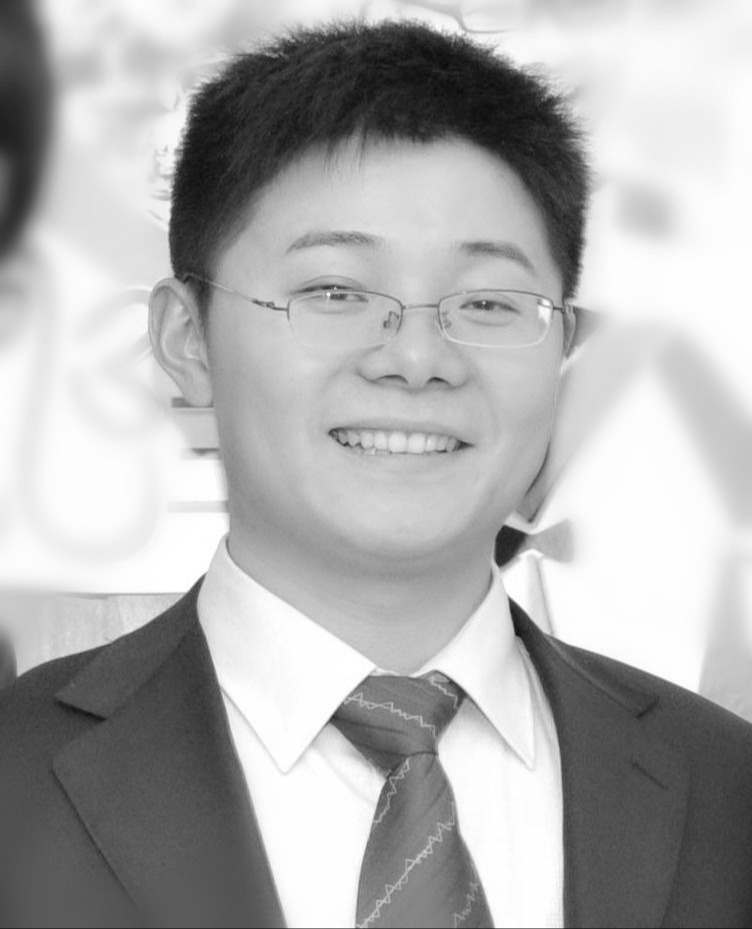}}]{Yong Li}
received the B.S. degree from Huazhong University of Science and Technology in 2007, and the M. S. and the Ph. D. degrees in Electrical Engineering from Tsinghua University, in 2009 and 2012, respectively. During 2012 and 2013, he was a Visiting Research Associate with Telekom Innovation Laboratories and Hong Kong University of Science and Technology respectively. During 2013 to 2014, he was a Visiting Scientist with the University of Miami. Currently, he is a Faculty Member of the Department of Electronic Engineering, Tsinghua University. His research interests are in the areas of big data, mobile computing, wireless communications and networking. Dr. Li has served as General Chair, TPC Chair, TPC Member for several international workshops and conferences, and he is on the editorial board of four international journals. His papers have total citations more than 7200 Google Scholar). Among them, ten are ESI Highly Cited Papers in Computer Science, and five receive conference Best Paper (run-up) Awards.
\end{IEEEbiography}

\end{document}